\documentclass[letterpaper]{article}
\usepackage{aaai}
\usepackage{fixbib}
\usepackage{times}
\usepackage{helvet}
\usepackage{courier}
\usepackage{graphicx}
\usepackage{enumerate}
\begin{document}
%
\title{Attendee-Sourcing: Exploring The Design Space of Community-Informed Conference Scheduling}
\author{
Anant Bhardwaj \\
anantb@csail.mit.edu\\
MIT CSAIL
\And
Juho Kim\\
juhokim@mit.edu\\
MIT CSAIL
\And
Steven Dow \\
spdow@cs.cmu.edu\\
Carnegie Mellon University
\AND
David Karger \\
karger@mit.edu\\
MIT CSAIL
\And
Sam Madden \\
madden@csail.mit.edu\\
MIT CSAIL
\And
Rob Miller\\
rcm@mit.edu\\
MIT CSAIL
\And
Haoqi Zhang \\
hq@northwestern.edu\\
Northwestern University
}
\maketitle
\begin{abstract}
\begin{quote}
Constructing a \emph{good conference schedule} for a large multi-track conference needs to take into account the preferences and constraints of organizers, authors, and attendees. Creating a schedule which has fewer conflicts for authors and attendees, and thematically coherent sessions is a challenging task. 

\emph{Cobi} introduced an alternative approach to conference scheduling by engaging the community to play an active role in the planning process. The current Cobi pipeline consists of \emph{committee-sourcing} and \emph{author-sourcing} to plan a conference schedule. We further explore the design space of \emph{community-sourcing} by introducing \emph{attendee-sourcing} -- a process that collects input from conference attendees and encodes them as preferences and constraints for creating sessions and schedule.  For CHI 2014, a large multi-track conference in human-computer interaction with more than 3,000 attendees and 1,000 authors, we collected attendees' preferences by making available all the accepted papers at the conference on a paper recommendation tool we built called \emph{Confer}, for a period of 45 days before announcing the conference program (sessions and schedule). We compare the preferences marked on Confer with the preferences collected from Cobi's author-sourcing approach. We show that attendee-sourcing can provide insights beyond what can be discovered by author-sourcing. For CHI 2014, the results show value in the method and attendees' participation. It produces data that provides more alternatives in scheduling and complements data collected from other methods for creating coherent sessions and reducing conflicts.
\end{quote}
\end{abstract}

\section{Introduction}
For large conferences with hundreds of papers and multiple parallel sessions, constructing a \emph{good} schedule is a difficult task. The difficulty of the task lies in how to create thematically coherent sessions and schedule them in such a way that it accommodates preferences and constraints of conference organizers, authors, and attendees. For logistic and pragmatic reasons, large-scale event planning typically becomes a responsibility of a few dedicated organizers ~\cite{EventPlanning}. Organizers have to often struggle to gather all the information necessary for an optimal planning. Even with sophisticated planning tools, organizers find it difficult to accommodate unforeseen issues and unvoiced opinions ~\cite{CoDesign}.

We observed the session creation process for ACM CHI, a multi-track large conference on Human-Computer Interaction (HCI). CHI 2013 received nearly 2,000 paper submissions and accepted almost 400. Until 2013, the schedule creation process at CHI involved two main stages. The conference organizers formed 80-minute sessions with 4-5 papers each, with 16 parallel sessions spanning four days. First, once papers are accepted, the technical program chairs and 15-25 committee members create small groups of papers to build a rough preliminary schedule. Apart from being dependent on individual domain expertise of the program chairs, this process is manual, paper-based, and time-consuming. In the second stage, the conference chairs refine this rough schedule to create the final program. They attempt to resolve conflicts, fix sessions with stray papers that do not fit, and generally strive to improve the program. The chairs make most changes via manual inspection. Despite organizers' best intentions and efforts, previous CHI programs often contained incoherent sessions, similarly themed sessions that run in parallel, and author/attendee-specific conflicts.

\emph {Cobi} introduced an alternative approach that draws on the people and expertise within the community, and embeds intelligence for resolving conflicts into a scheduling interface. Once the organizing committee determines the accepted papers, Cobi invites committee members to group papers in their specific areas of expertise~\cite{FrenzyCHI2014}. Once committe members group papers, Cobi invites the authors of the accepted papers to identify papers that would fit well in a session with their own, and which papers would they like to see.  The authors are presented with 20 papers most similar to their paper, and two questions: ``how relevant is each paper (i.e., should it be in the same session as the author's paper)?'' and ``is the paper interesting (i.e., would the author like to see this paper's presentation)?''. The relevance feedback from authors provides fine-grained information about which papers should appear in the same session. The interest feedback helps in creating sessions and schedule that allows authors to see as many papers as they are interested in without conflicting with other sessions at the same time. These inputs are encoded as preferences and constraints, and presented in Cobi's scheduling tool to help conference organizers create sessions, resolve conflicts, and build the schedule ~\cite{CobiUIST2013}. Cobi's author-sourcing data helps conference organizers make an informed decision about which papers to group together, and resolve conflicts by highlighting author-specific conflicts (papers an author would be interested in seeing are scheduled at the same time). \emph{Readers should note that for the rest of the paper, \emph{``author-sourcing''} refers to Cobi's author-sourcing (a specific instance of collecting inputs from authors for conference scheduling).}

Despite the many benefits, we identify the following limitations in Cobi's author-sourcing approach:
\begin{enumerate}
\item \textbf{P1:} Authors can't indicate any paper outside the presented list (20 papers) as ``similar to their paper'', or ``would like to see''. The conflicts can't be fully resolved because there could be papers that authors may want to see but they have no way to express in their author-sourcing response.
\item \textbf{P2:} Attendees haven't been reached out. This is important because for a truly community-driven conference scheduling, all attendees' inputs are crucial -- especially because committee members and authors are small in number compared to the entire community. It is not possible to capture and resolve non-author-specific conflicts through author-sourcing.
\end{enumerate}

\begin{figure}[!h]
\centering
\includegraphics[width=0.9\columnwidth]{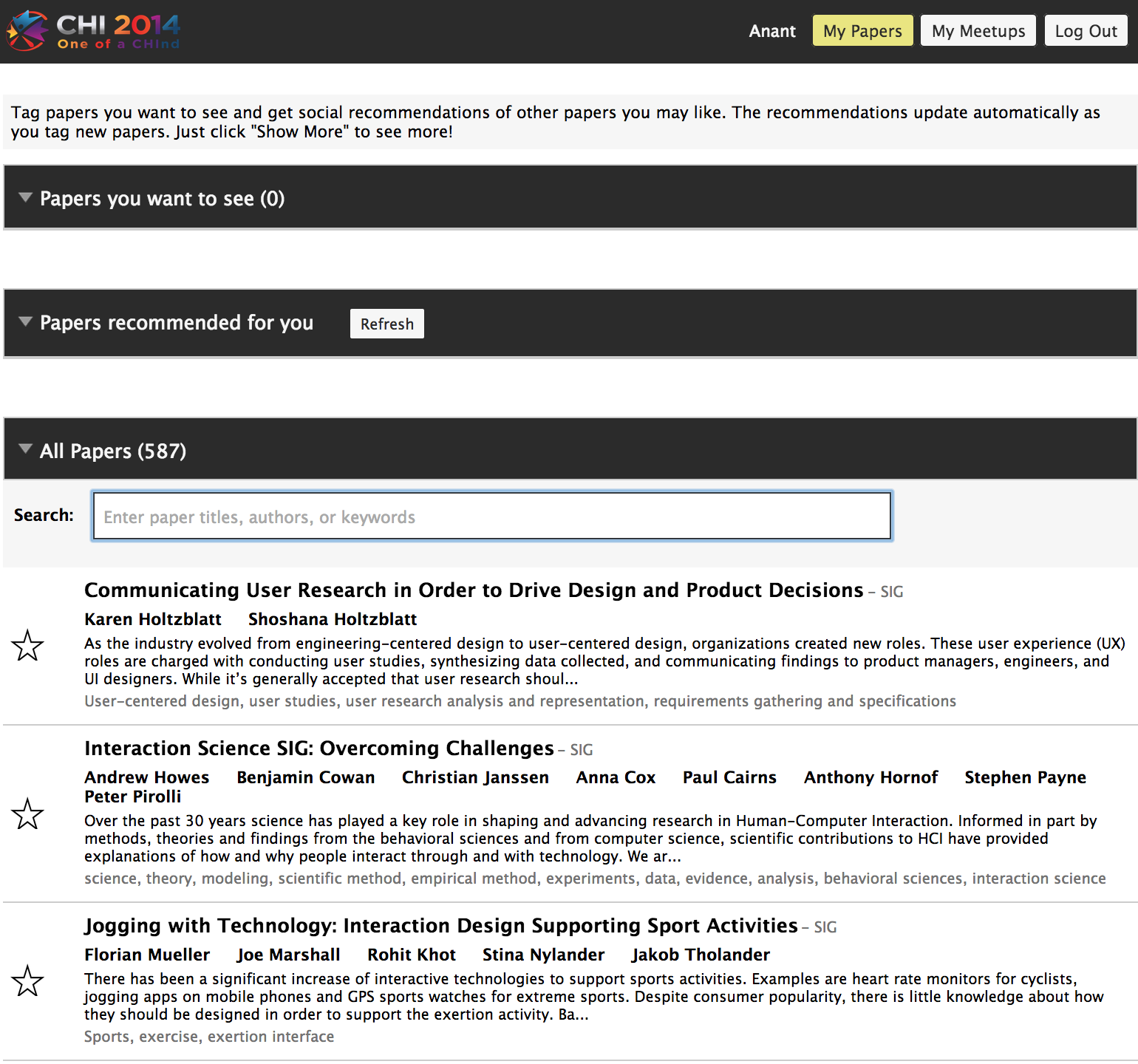}
\caption{\emph{For CHI 2014, all the accepted papers were released on Confer 4 months before the conference (45 days before releasing the schedule).}}
\label{all-accepted-papers}
\end{figure}

In this paper, we explore the design space of \emph{community-sourcing} and propose \emph{attendee-sourcing} -- a process that collects input from conference attendees and encodes them as preferences and constraints for creating sessions and schedule. We give attendees an interface that allows them to discover and explore the entire set of papers without restriction. Attendees can mark as many papers as they like in their preference list. For CHI 2014, we collected attendees' preferences by putting all the papers accepted at CHI 2014 into a paper-recommendation tool we built called Confer ~\cite{Confer} (Figure~\ref{all-accepted-papers}), for a period of 45 days before releasing the conference program (sessions and schedule). We find that attendees are often interested in exploring award/honorable mention papers, papers from a particular institution/organization, papers by a particular author, papers related to a particular topic, etc. In the process of exploration, they find interesting papers which they add to their preference list. We believe that with attendee-sourcing we can collect rich preference data as by-products of \emph{natural} exploration by attendees. The Confer interface and underlying recommendation engine are designed to support the exploratory behavior and keep the attendees engaged by showing them papers they are most likely to find interesting.

We compare the data collected from \emph{attendee-sourcing} with the data collected from Cobi's \emph{author-sourcing}. Our analysis suggests that attendee-sourcing can provide affinities beyond what can be discovered by author-sourcing. We find that both approaches provide unique insights difficult to capture by the other. We conclude that the two approaches complement each other and provide alternatives for creating coherent sessions and reducing conflicts.

The paper is organized as follows. First, we describe related work in the community-sourcing design space. Next, we discuss the design motivation for \emph{attendee-sourcing} and compare it with author-sourcing. We then discuss the data collected via the attendee-sourcing interface for CHI 2014 and compare it with the author-sourcing data. In the following section, we analyze the differences and present a number of ways in which they differ. Before conclusion, we discuss the implications of these differences and show how these two sources can complement each other.

\section{Related Work}
Our work explores the incentives, methods, and interfaces for collecting and incorporating preferences and constraints from a large, heterogeneous community. We also explore how to design community-sourcing methods for members with different roles and incentives, who can produce different data.

The most recent work in this area is Cobi~\cite{CobiUIST2013} which initially involves committee members to group papers and then invites the authors of the accepted papers to identify papers that would fit well in a session with their own, and which papers would they like to see. Cobi makes a design choice of giving the authors a list of 20 similar papers to choose from. While simplifying the task for authors, they can't indicate any paper outside the presented list (20 papers) as ``similar to their paper'', or ``would like to see''. We argue that it is possible to miss many paper-similarities because of such design.

A number of recent systems explore the use of crowds for making and executing plans ~\cite{CollaborativePlanning}, ~\cite{Turkomatic}, ~\cite{HCT}, and ~\cite{EmailVallet}. Prior work on community-sourcing has shown that community-sourcing can successfully elicit high-quality expert work from specific communities. ~\cite{CommunitySourcing} show that students can grade exams with 2\% higher accuracy (at the same price) or at 33\% lower cost (at equivalent accuracy) than traditional single-expert grading. ~\cite{CrowdDiversity} explore how expert input can be combined with input from members of the general population. This provides a motivation for attendee-sourcing. We consider attendees as general population, and the data collected from attendee-sourcing can be combined with expert inputs from conference organizers for session creation and schedule planning.

Another thread of research explored automatically generating an optimal schedule. In the context of conference scheduling, ~\cite{AutomaticScheduling} introduced formulations for maximizing the number of talks of interest attendees can attend. However, ~\cite{CobiUIST2013} identified that while automated scheduling is appropriate when the parameters and constraints of the optimization problem are well-specified, interviews with past CHI organizers show that they attempt to tackle soft constraints and other tacit considerations. We discuss how attendee-sourcing data could be useful for automatically generating  a draft schedule, which the organizers can refine with their domain expertise and subjective considerations.

Confex ~\cite{Confex} is a commercial solution for conference management, which includes tools for scheduling. To the best of our knowledge, no existing conference management solution incorporates community input directly in the scheduling process. Conference Navigator ~\cite{ConferenceNavigator} and Conferator ~\cite{ConferenceAnatomy} provide a similar set of features to Confer, including bookmarking papers of interest and building personalized schedules with paper recommendations. These systems, however, have not used the attendees' data outside of the tool itself. Our attendee-sourcing introduces a novel mechanism that uses the by-product of attendees' paper exploration in the scheduling process. Our method can even use data generated by these other tools for improving the schedule.

Collaborative filtering ~\cite{CollaborativeFiltering} has proven to be valuable for recommending items in many different domains. ~\cite{CitationRecommendation} have explored the use of collaborative filtering to recommend research papers, using the citation web between papers to create the ratings matrix. We construct the ratings matrix from attendees' bookmarks to recommend papers.

\section{Author-Sourcing vs. Attendee-Sourcing}
In this section, we explain the \emph{author-sourcing} process used by Cobi, and the proposed \emph{attendee-sourcing} process. We compare the two w.r.t the interface design, the design rationale, incentives for participants (authors vs. attendees), nature of the data collected, and how the data can be used for creating sessions and planning the schedule.
\begin{table*}[t]
  \centering
\begin{tabular}{ | p{5.5cm} || p{5cm} | p{5.5cm} | }
\hline
\textbf{Metric} & \textbf{Author-Sourcing} & \textbf{Attendee-Sourcing} \\ \hline
\hline
\# of participants & 634 & 347 \\ 
\hline
\# unique papers marked & 535 (91.14\% of all the papers) & 587 (98.98\% of all the papers) \\ 
\hline
\# preferences marked & 3,869 & 7,228 \\
\hline
\# preferences per participant & \emph{mean:} 6.1 \emph{median:} 5 \emph{std-dev:} 3.86  & \emph{mean:} 20.83 \emph{median:} 14 \emph{std-dev:} 21.47 \\
\hline
\# of participants' preferences per paper & \emph{mean:} 7.23 \emph{median:} 6 \emph{std-dev:} 4.7  & \emph{mean:} 12.44 \emph{median:} 10 \emph{std-dev:} 7.98 \\
\hline
\end{tabular}
\caption{\emph{A summary of participation and coverage statistics for author-sourcing and attendee-sourcing deployed to CHI 2014}}
\end{table*}

\subsection{Author-Sourcing}
\begin{figure}[!h]
\centering
\includegraphics[width=0.9\columnwidth]{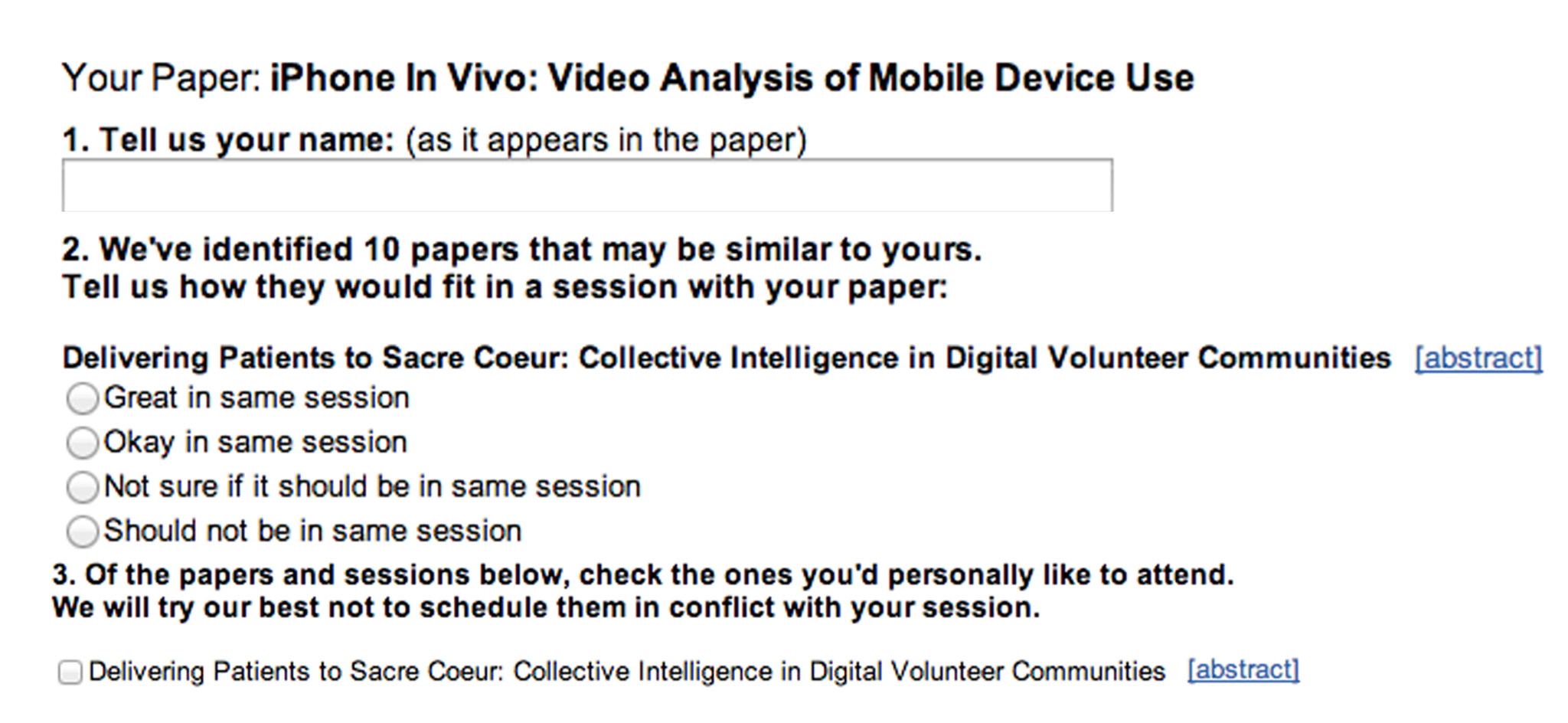}
\caption{\emph{Authors are presented with a custom list of 10-20 papers and asked to judge which are related to their paper or of interest to them.}}
\label{author-sourcing}
\end{figure}

To prepare initial paper similarity, Cobi uses groups generated from committee-sourcing along with TF-IDF scores on paper titles and abstracts to generate 20 most similar papers for every paper ~\cite{CommunityClusteringHCOMP13}. In the author-sourcing stage, Cobi presents the 20 papers to all paper authors (Figure~\ref{author-sourcing}). The interface presents only the first 10 initially, and provides a \emph{``show more''} link to see the next 10.  The authors are presented with two questions: 1. how relevant is each paper (i.e., should it be in the same session as the author's paper); and 2. is the paper interesting (i.e., would the author like to see this paper's presentation). The relevance feedback from authors provides fine-grained information about which papers should appear in the same session. The goal of interest feedback is to create sessions and schedule in such a way that it allows authors to see as many papers as they are interested in without conflicting with other sessions at the same time.
\begin{enumerate}[I]
\item \textbf{Interface \& Design Rationale:} Figure~\ref{author-sourcing} shows the author-sourcing interface used by Cobi to get relevance and preference feedback from authors. The interface presents users with a list of papers which committee-sourcing + TF-IDF identified as similar. The authors are most likely the best people to judge whether the other papers in the same group are related to their paper. The choice of keeping the list small is critical because otherwise authors may get overwhelmed by a huge number of potentially unrelated papers. It is likely that the participation will reduce significantly if all the papers (587 at CHI 2014) are presented to the authors to choose from. However, this introduces a problem - authors can't indicate any paper outside the presented list (20 papers) as ``similar to their paper'', or ``would like to see''. The conflicts can't be fully resolved because there could be papers that authors may want to see but they have no way to indicate those in their author-sourcing response.

\item \textbf{Incentives for Participants:} The authors are motivated to participate in the process so that their paper may end up in a session with relevant papers. The process also provides them with an advance preview of \emph{some} of the accepted papers before the program is announced. For CHI 2014, author-sourcing involved 634 authors - they marked 3869 preferences in total \emph{(mean: 6.1, median: 5, std-dev: 3.86, min: 1, max: 40)}. Some authors got multiple papers accepted at the conference -- they participated for all their papers causing  $max > 20$. It covered 535 papers (91.14\% of all the papers). On an average, a paper was in the preference list of 7.23 authors \emph{(mean: 7.23, median: 6, std-dev: 4.7, min: 1, max: 32)}.

\item \textbf{Nature of the data collected \& potential use:} Two important types of data are collected: the relevance data which is an important input for making sessions that are coherent, and preference data which can be used to reduce author-specific conflicts (conference organizers will try not to schedule two papers an author wants to see at the same time). A limitation of this approach is that non-author attendees haven't been reached out. This is important because for a truly community-driven conference scheduling, all attendees' inputs are crucial -- especially because committee members and authors are small in number compared to the entire community. We later describe how members with different roles and different incentives produce different data.
\end{enumerate}
\subsection{Attendee-Sourcing}
We use the \emph{Confer} interface (Figure~\ref{all-accepted-papers}) for attendee-sourcing. Confer is a tool that lets conference attendees mark papers they are interested in to get social recommendations and a personalized schedule.

\begin{figure}[!h]
\centering
\includegraphics[width=0.9\columnwidth]{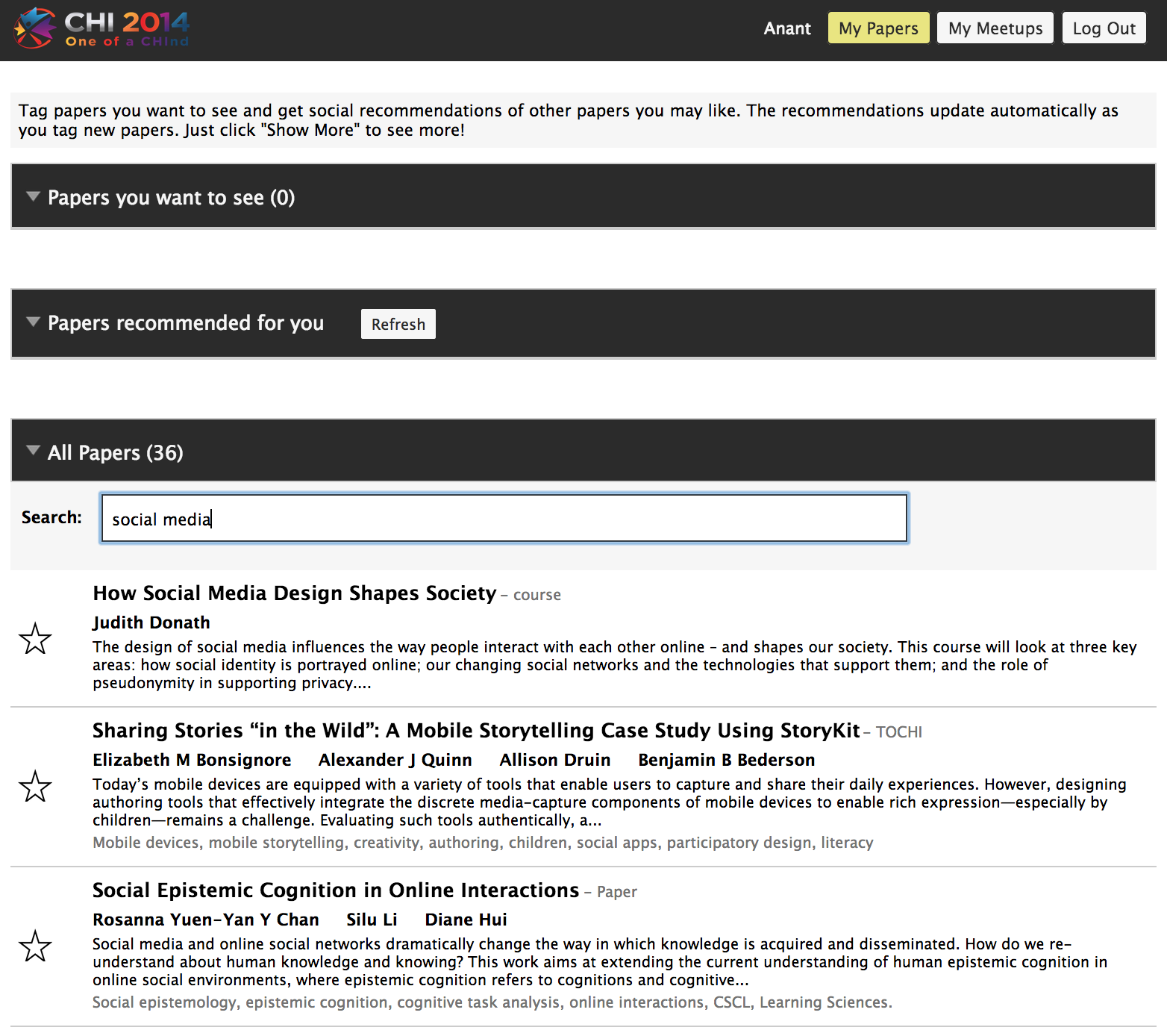}
\caption{\emph{Attendees can search for papers using keywords, author names, affiliation, etc.}}
\label{confer-search}
\end{figure}

For CHI 2014, the program chairs released the accepted papers (Figure~\ref{all-accepted-papers}) on Confer 4 months in advance (45 days before releasing the sessions and schedule). The CHI 2014 organizing committee also ensured that the first list of papers was \emph{only} available on Confer. It was simply a list of papers with no sessions and schedule information. The chairs made an official announcement on Twitter (Figure~\ref{chi-announcement-2}) that attendees can explore and mark the papers they want to see at the conference. They also announced that the preferences collected from Confer would be used for optimal session planning (Figure~\ref{chi-announcement-1}). Conference attendees marked papers they want to see at the conference (Figure~\ref{attendee-response}) by either searching (Figure~\ref{confer-search}) or using recommendations (Figure~\ref{confer-recommendations}).

\begin{enumerate}[I]
\item \textbf{Interface \& Design Rationale:} The Confer interface is designed to give attendees a way to discover and explore without any restriction. It is designed to support {natural exploration} by attendees. To support exploration, Confer uses collaborative filtering ~\cite{CollaborativeFiltering} based social recommendations. In the beginning, when there aren't enough preferences to support collaborative filtering, the system automatically falls back to content (TF-IDF) based recommendations. This ensures that attendees always get some relevant recommendations when they interact with the system. The design goal is to keep the attendees engaged by showing them papers they are most likely to find interesting.
 
\begin{figure}[!h]
\centering
\includegraphics[width=0.9\columnwidth]{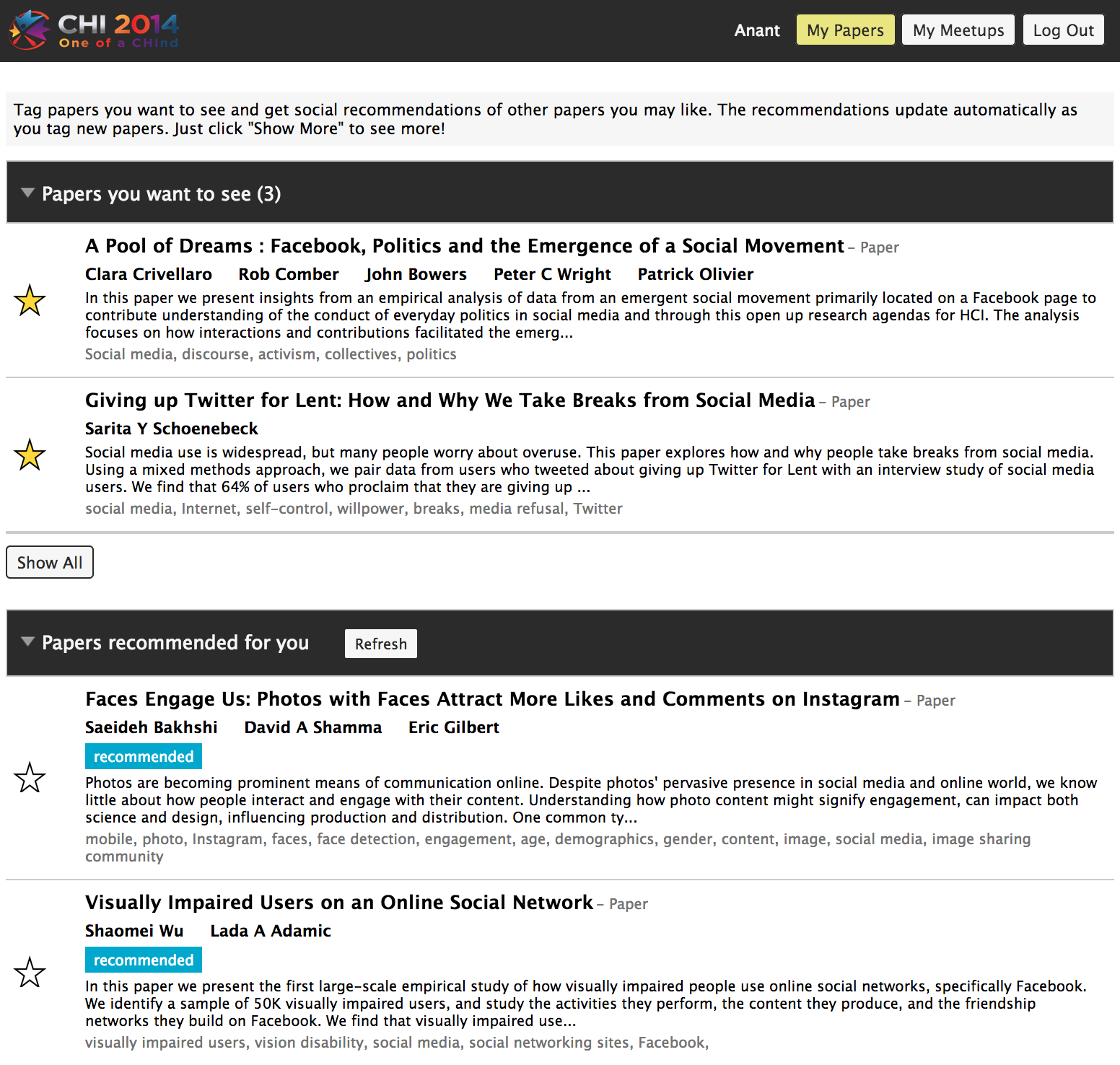}
\caption{\emph{Attendees are presented with a set of recommendations based on their current selection to help them discover relevant papers.}}
\label{confer-recommendations}
\end{figure}
\item \textbf{Incentives for Participants:} The primary motivation for attendees is to get an early preview of all the accepted papers which they can explore and interact with. From our logs we find that attendees search for papers from a particular institution/organization, papers by a particular author, papers related to a particular topic, etc. They spend time exploring different papers by clicking the recommendations we show. While exploring, they add interesting papers to their preference list. The Confer interface and underlying recommendation engine are designed to support the exploratory behavior and keep the attendees engaged by showing them papers they are most likely to find interesting. To encourage participation, after releasing the accepted papers on Confer, CHI officially announced that attendees should explore CHI 2014 papers on Confer (Figure~\ref{chi-announcement-2}). CHI also put Confer links in all the official announcements for honorable mention and best paper awards. 

\begin{figure}[!h]
\centering
\includegraphics[width=0.9\columnwidth]{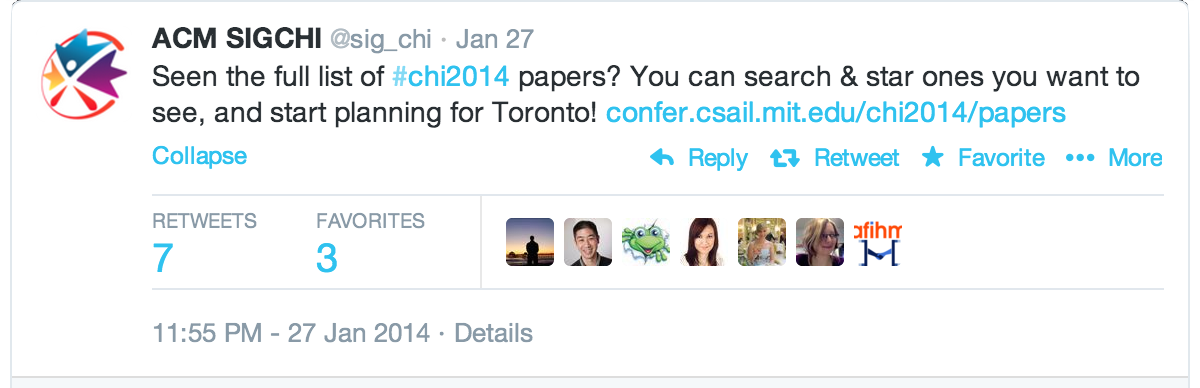}
\caption{\emph{CHI announcement on Twitter: ``Attendees can explore CHI 2014 papers on Confer''}}
\label{chi-announcement-2}
\end{figure}

Attendees also have an inherent motivation that they want a schedule which takes into account their preferences (Figure~\ref{chi-announcement-1}) so that they can see all the papers they want to see without any conflict. A conflict is when two papers that an attendee wants to see are scheduled at the same time. To motivate attendees, CHI officially made an announcement that the data collected from Confer would be used to determine the optimal sessions.

\begin{figure}[!h]
\centering
\includegraphics[width=0.9\columnwidth]{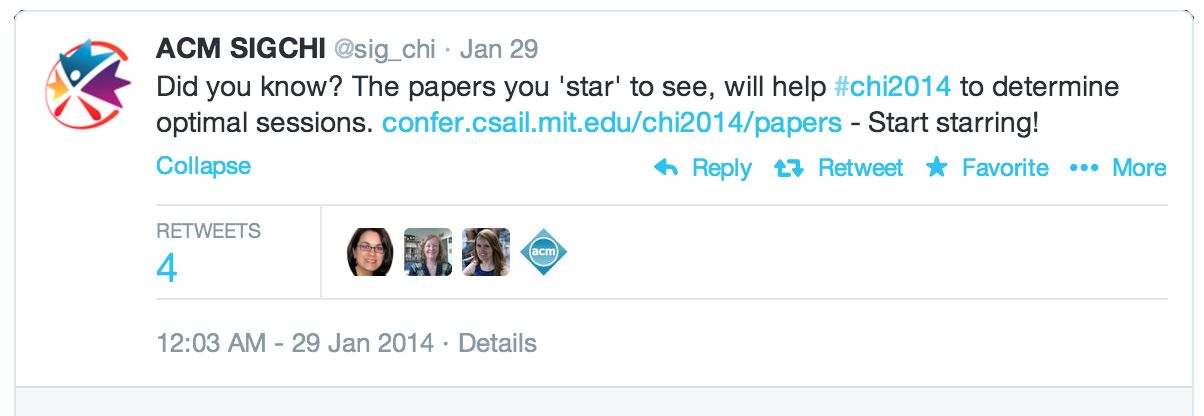}
\caption{\emph{CHI announcement on Twitter: ``attendees' preference would be used to create optimal sessions''}}
\label{chi-announcement-1}
\end{figure}

We saw a good response from attendees (Figure~\ref{attendee-response}). For CHI 2014, attendee-sourcing attracted 347 attendees - they marked 7228 preferences in total \emph{(mean: 20.83, median: 14, std-dev: 21.47, min: 1, max: 156)}. It covered 581 papers (98.98\% of all the accepted papers). On an average, each paper was in the preference list of 12.44 attendees \emph{(mean: 12.44, median: 10, std-dev: 7.98, min: 1, max: 56)}.

\begin{figure}[!h]
\centering
\includegraphics[width=0.9\columnwidth]{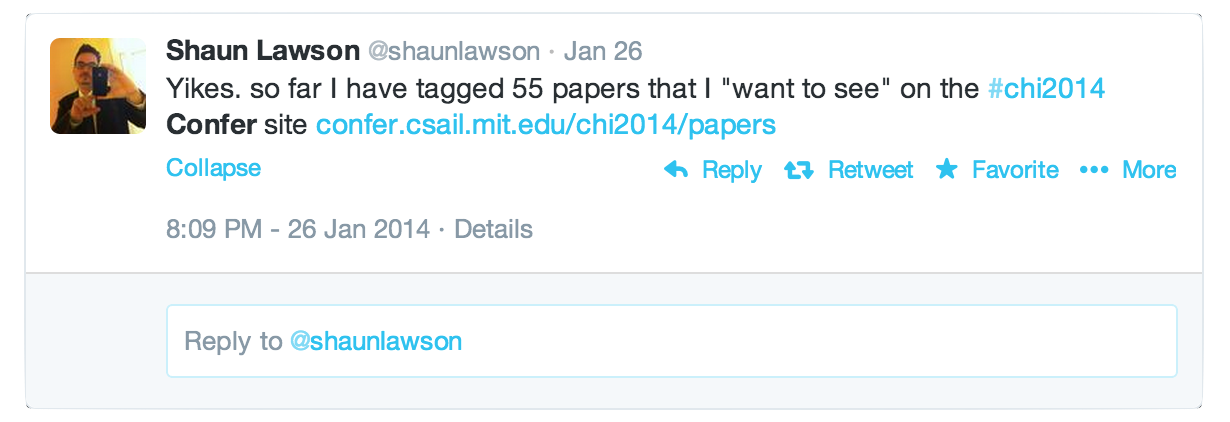}
\caption{\emph{An attendee tweets about his ``wants to see'' list for CHI 2014}}
\label{attendee-response}
\end{figure}
\item \textbf{Nature of the data collected \& potential use:} The data collected from \emph{attendee-sourcing} is attendees' preferences resulting from a natural exploration. The process involves all the attendees -- authors as well as non-authors. Unlike author-sourcing where authors can't indicate any paper outside the presented list (20 papers), attendees can add any paper from the conference to their ``want to see'' list. With attendee-sourcing, it is possible to reach out to all members of the community. We saw an active participation for CHI 2014. If more attendees participate, this could help organizers further improve the schedule and encourage even more attendees to explore the schedule with additional incentives. 

There is a flip side too though -- the preferences marked by attendees are susceptible to external influence such as Confer recommendations, social trends (links shared on social media by friends), CHI award/honorable mention tweets, etc. Also, the preference signal might require additional interpretation for it to be used to infer relevance between papers.

We present the following hypotheses for the potential use of \emph{attendee-sourcing} data:

\begin{enumerate}
\item \textbf{H1:} Attendee-sourcing provides paper affinities beyond what can be discovered via committee-sourcing and author-sourcing.

\item \textbf{H2:} Attendee-sourcing provides more alternatives in scheduling and can complement author-sourcing for creating coherent sessions and reducing conflicts.
\end{enumerate}
\end{enumerate}

In the next section, we analyze the data collected from both methods.
\section{Data Analysis}
In this section we present a comparative analysis for \emph{attendee-sourcing} and \emph{author-sourcing} data collected for CHI 2014. 
\begin{enumerate}
\item \textbf{Participation}: Author-sourcing involved 634 authors - they marked 3,869 preferences in total (mean: 6.1, median: 5, std-dev: 3.86, min: 1, max: 40). Some authors got their multiple papers accepted at the conference -- they participated for all their papers causing  $max > 20$. Attendee-sourcing attracted 347 attendees - they marked 7,228 preferences in total (mean: 20.83, median: 14, std-dev: 21.47, min: 1, max: 156).

This data clearly shows that attendees want to see more papers than what they are able to indicate in their author-sourcing response. This difference is significant -- on an average, an attendee wanted to see 20.83 papers but s/he indicated only 6.13 in the author-sourcing response. This suggests that attendee-sourcing can capture a lot of preference data not captured by author-sourcing.

Although author-sourcing has better participation (\# of participants), attendee-sourcing is able to generate more preference data points with a smaller number of attendees. Also, we find that after a certain threshold, the similarity structure doesn't change much even if we get data from more attendees. We however agree that there is a need to find ways to incentivize attendees so that the \# participation count can be increased.

\item \textbf{Coverage}: Author-sourcing responses covered 535 papers (91.14\% of all the accepted papers). On an average, each paper was in the preference list of 7.23 authors (mean: 7.23, median: 6, std-dev: 4.7, min: 1, max: 32). Attendee-sourcing responses covered 581 papers (98.98\% of all the accepted papers). On an average, each paper was in the preference list of 12.44 attendees (mean: 12.44, median: 10, std-dev: 7.98, min: 1, max: 56)

The statistics above shows that attendee-sourcing provides more coverage (\# of papers for which preference data is available) as well as more \# of preference data points for each paper.

\item \textbf{Capturing affinity}: We define affinity score between a pair of papers as \# of people who want to see both papers. Readers might argue about possible biases, which we discuss later. As we saw, for attendee-sourcing the expected number of attendees who want to see a paper is 12.44; if a pair of papers is in the preference list of more than 10 attendees, we consider it a strong signal for relevance. Similarly, for author-sourcing the expected number of authors who want to see a paper is: 7.23; if a pair of papers is in the preference list of more than 5 authors, we consider it a strong signal for relevance.

We compare attendee-sourcing and author-sourcing based on affinity score and discover the following:
\begin{enumerate}
\item \emph{We didn't find any pair for which attendee-sourcing reported a zero affinity and author-sourcing reported a strong ($>5$) affinity}. This shows that attendee-sourcing captures everything that author-sourcing can discover. Also, there were only 28 pairs for which attendee-sourcing detected a weak affinity ($<=10$) and author-sourcing detected a strong ($>5$) affinity. To a large extent attendee-sourcing affinities form a super set of author-sourcing affinities.

\item There are 402 pairs, for which the difference between the affinity scores reported by attendee-sourcing and author-sourcing is $>10$. Moreover, there are 150 pairs for which author-sourcing reports a \emph{zero} affinity and attendee-sourcing reports a strong ($>10$) affinity. In the following paragraphs, we present our reasoning for this difference.

\end{enumerate}

In author-sourcing, authors select their preferences from a set of pre-computed potentially related papers, so the similarities computed from author-sourcing is just a validation and refinement of the pre-computed similarities. Author-sourcing can't discover a new similarity by itself. On the other hand, attendee-sourcing preferences are by-products of a natural navigation and thus attendee-sourcing can discover new similarities.

It is fair to assume that if enough people mark a pair of papers together in their preference list, there is a high likelihood that the pair is related. As we can see, there exists a large number of pairs for which the affinity data differs, especially the 150 pairs for which author-sourcing fails to capture any affinity at all. We take one example pair to understand this -- let's consider the following two papers:
\begin{enumerate}
\item \emph{Combining Body Pose, Gaze, and Gesture to Determine Intention to Interact in Vision-Based Interfaces}: The paper describes an algorithm that combines facial features, body pose and motion to approximate a user's intention to interact with a gesture based system. The algorithm can be used to determine when to pay attention to a user's actions and when to ignore them. \textbf{Keywords:} free-space interaction, vision-based input, user engagement, input segmentation, learned models

\item \emph{Vulture: A Mid-Air Word-Gesture Keyboard}: This paper is about a word-gesture keyboard that lets users draw the shape of a word in mid-air. It extends touch based word-gesture algorithms to work in mid-air. It projects users' movement onto the display, and uses pinch as a word delimiter. \textbf{Keywords:} word-gesture keyboard, shape writing, text entry, mid-air interaction, in-air interaction, freehand interaction
\end{enumerate}

Committee-sourcing captures mostly the salient signals (information based on keywords, title, abstract, etc.) of the papers to form initial groups. These signals may not be always enough to conclude whether two papers are related. For the above two papers, it's difficult to discover that they are closely related without having a deeper understanding of the algorithms they describe.  Also, these two papers have low TF-IDF similarity based on textual information. Since author-sourcing takes the similarity input from the previous stage (committee-sourcing + TF-IDF), it will miss these similarities. In general, the main limitation of these methods is that they can't capture all the dimensions of similarity between papers. If a paper spans multiple sub-fields or domains, it might miss some of those. Attendees, on the other hand, can capture this collectively, because attendees bookmark papers based on their own interest and expertise. When we combine diverse data points from individuals, such multi-faceted nature of papers can be captured. In summary, \emph{author-sourcing focuses on validating similarity, while attendee-sourcing can discover new similarity}.

It is important to note that these similarities are important from the perspective of creating coherent sessions and minimizing conflicts in the schedule.
\end{enumerate}

Now we show how attendee-sourcing data can be used for creating coherent sessions and reducing conflicts.  In particular, attendee-sourcing can be used to:
\begin{enumerate}
\item \emph{Create Coherent Sessions.} If many attendees want to see a pair of papers together, then the two papers can be categorized as related. Ideally, it would make sense to put a set of papers a lot of attendees want to see together in the same session. Sessions can be generated by creating clusters based on affinity scores computed from attendee-sourcing preferences.

\item \emph{Reduce Conflicts.} This is straight-forward -- don't schedule two papers in different sessions at the same time if many attendees want to see both of them. 

\item \emph{Measure Paper Popularity.} Determining paper popularity in advance is useful for making strategic decisions on the organizers' side. For example, when assigning rooms, organizers can put popular papers in bigger rooms, and less popular papers in smaller rooms. More accurate popularity estimates can avoid situations like Figure~\ref{crowded-room}, which shows a session from CHI 2013 when a paper a lot of attendees wanted to see was assigned a smaller room. An alternative strategy for scheduling would be as follows: assign a paper very few people want to see in a session with more popular papers, rather than clustering all the less popular papers in the same session. This can help conference organizers maintain popularity balance in the schedule.

\begin{figure}[!h]
\centering
\includegraphics[width=0.9\columnwidth]{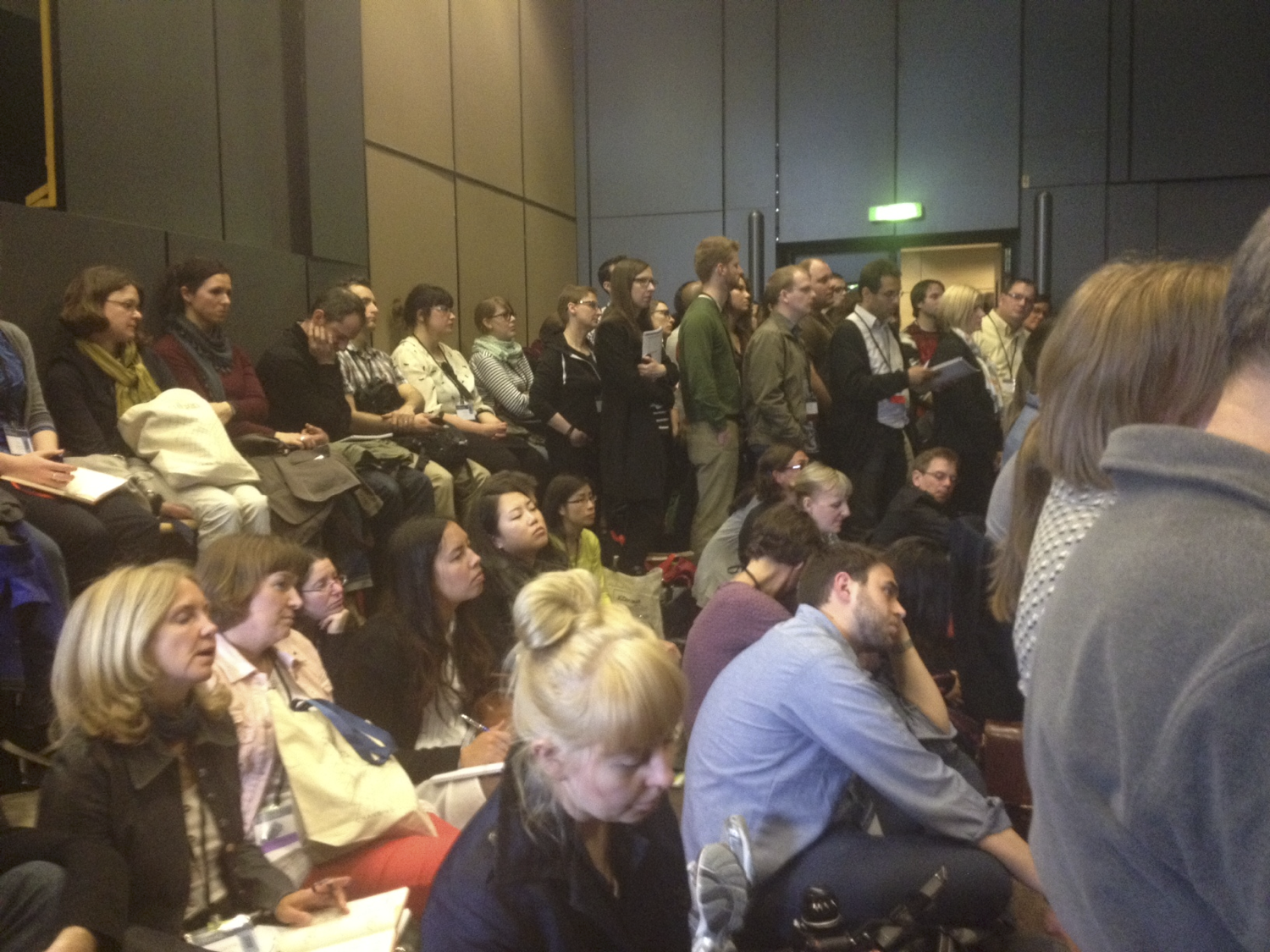}
\caption{\emph{CHI 2013: A popular paper scheduled in a small room (CHI 2013 had only author-sourcing)}}
\label{crowded-room}
\end{figure}
\end{enumerate}

\section{Discussion}
One interesting aspect of this paper is to look at the layers of community-sourcing inside a complex community. Readers might find it interesting how vastly different design approaches have been used to collect input from authors and attendees, respectively. One must note that authors are also attendees. They are not mutually exclusive groups but attendees are just larger. Our experience shows that many authors also participated in attendee-sourcing, perhaps because authors have different incentives as being attendees.

Now we present a discussion w.r.t the two hypotheses we had mentioned earlier - \textbf{H1:} Attendee-sourcing provides paper affinities beyond what can be discovered via committee-sourcing and author-sourcing, and \textbf{H2:} Attendee-sourcing provides more alternatives in scheduling and can complement author-sourcing for creating coherent sessions and reducing conflicts.
\begin{enumerate}
\item \emph{Author's incentive is to have related papers in the same session as his/her paper, Attendee's incentive is to find all the papers s/he wants to see}: Different incentives and roles produce different data. From author-sourcing we get which papers authors want to see in the same session as their paper, a valuable input for creating coherent sessions. From attendee-sourcing, we get all the papers one wants to see at the conference, which gives us rich data for detecting cross-session conflicts, balancing sessions, finding rooms of right size, etc. Combining data from both the sources gives conference organizers a way to make an informed decision about many different aspects of conference planning.

\item \emph{Authors validate similarity, Attendees discover similarity}: Different interfaces produce different data. The author-sourcing interface helps validate affinities computed from the previous stage (committee-sourcing + TF-IDF). Authors are presented with a static set of potentially similar papers. Authors are the best people to judge which papers are related to their paper. On the other hand, the attendee-sourcing can discover new similarities. The interface is designed for a natural exploration. Confer's recommendations are continuously adapting based on what one marks in his/her preference list and the changes happening in the network. If a large number of attendees mark a set of papers together, there is a high likelihood that the papers in the set are related. Attendee-sourcing can discover new similarities.

\item \emph{Author-Sourcing gives Precision, Attendee-Sourcing gives Recall}: Different methods produce data of different nature. Author-sourcing can best capture paper affinity because the authors are the best people to judge which papers are related to their paper.  Author-sourcing presents the authors a set of 20 papers likely to be similar and asks them to mark papers that are related to their paper. Author's feedback of paper similarity can be regarded as a strong signal of similarity. The affinity computed from author-sourcing data is guaranteed to have a \emph{high precision}. On the flip side, there is no way authors can indicate papers outside the presented list as similar to their paper. This adversely affects the \emph{recall}. We earlier saw that for 150 pairs of papers, attendee-sourcing produced strong affinity but author-sourcing produced \emph{zero} affinity. Affinities computed from attendee-sourcing are likely to have lower precision because the process is susceptible to external influences such as social trends (links shared on social media by friends), CHI award/honorable mention tweets, etc. Combining both will result in a high quality similarity detection. Initially we can detect similar papers from attendee-sourcing and then use author-sourcing to further refine it.
 \end {enumerate}

\section{Implications \& Future Work}
Our analysis suggests the following implications and future work:
\begin{enumerate}
\item \textbf{Engage All Community Members:} We have demonstrated that data collected from one source can't capture everything. Members with different roles (authors, attendees, committee members, etc.) and different incentives produce different data. Also, members produce different data depending upon the interface they are presented with.

\item \textbf{Author-Sourcing Provides Valuable Data:} Author-sourcing attracts a significant participation -- for CHI 2014, authors of 91.14\% of all papers participated in the process. Authors have the inherent motivation to see their paper in a session of related papers. Data collected from author-sourcing reinforces and refines the affinity data computed through committee-sourcing. The refined affinity is very useful for grouping papers into coherent sessions. While the affinity data can also be collected from attendee-sourcing, it is susceptible to noise because attendee-souring preferences is susceptible to external influence. This makes author-sourcing very valuable.

\item \textbf{Attendee-Sourcing Provides Valuable Data:} We have demonstrated that attendee-sourcing can provide affinities beyond what can be discovered via author-sourcing. We saw a large number of pairwise affinities that attendee-sourcing could detect but weren't captured in author-sourcing. While we see attendees' participation (347) much lower than authors' participation (634) in the process, the former could still produce a lot more data-points (7,228 total preferences) than the latter (3,869 preferences). The results clearly show value in the method and attendees' participation. It produces data that could provide more alternatives in scheduling. To encourage more participation from attendees, conference organizers may attempt to recognize attendees' contributions explicitly, via social or even financial methods. The incentive for conference organizers is that they can improve the overall quality of the schedule and engage more community members to have ownership in the conference.

\item \textbf{Automated Session Creation:} In the future, we plan to explore automatically creating sessions and schedule to further reduce the burden on conference organizers. One possible approach to automatically generating a schedule is to consider it as an optimization problem with two optimization goals - low conflict, and high session coherence. The automatic schedule generator can use both author-sourcing as well as attendee-sourcing data to meet the optimization goals. The algorithmically generated schedule can be given as an input to Cobi's scheduling tool where conference organizers can interactively make changes, and further refine it.

\item \textbf{Beyond Sessions and Schedule:} Community-sourced data is useful not just for creating sessions and schedule but for making strategic decisions. The data can be used to understand the community structure -- specifically the areas of interest and the connections between them. Attendee-sourcing provides rich preference data that can be used to detect networks within the community and understand the connections within and outside the network. A good understanding of community network can be useful in discovering new areas of research and practice, new methodologies, and emerging application areas.
\end{enumerate}
\section{Conclusion}
This paper explores the incentives, methods, and interfaces for collecting and incorporating preferences and constraints from large numbers of individuals within an academic community for conference scheduling. We explore the design space of how members with different roles and incentives can produce complementary data.

For CHI 2014, we collected attendees' preferences by putting all the papers accepted at CHI 2014 into a paper recommendation tool we built called Confer for a period of 45 days before releasing the conference program. We compared the preferences marked on Confer with the preferences collected from Cobi's author-sourcing approach. The results show that attendee-sourcing can provide insights beyond what can be discovered by author-sourcing. We discuss the underlying design dimensions which cause attendee-sourcing to produce data of very different nature than author-sourcing. We show that the data collected from different approaches vary in nature and complement each other because of differences in the roles, incentives, and interfaces presented to the participants.

We believe the methods and insights presented in this paper generalize not just to scheduling other academic conferences, but also to other tasks that require collecting inputs from groups with different roles and incentives within a community.
\bibliography{references.bib}

\begin{thebibliography}{}

\bibitem[\protect\citeauthoryear{Allen}{2000}]{EventPlanning}
Allen, J.
\newblock 2000.
\newblock {\em Event Planning: The Ultimate Guide To Successful Meetings,
  Corporate Events, Fundraising Galas, Conferences, Conventions, Incentives,
  And Other Special Events}.
\newblock Toronto: John Wiley \& Sons Canada.

\bibitem[\protect\citeauthoryear{Andr{\'e} \bgroup et al\mbox.\egroup
  }{2013}]{CommunityClusteringHCOMP13}
Andr{\'e}, P.; Zhang, H.; Kim, J.; Chilton, L.~B.; Dow, S.~P.; and Miller,
  R.~C.
\newblock 2013.
\newblock Community clustering: Leveraging an academic crowd to form coherent
  conference sessions.
\newblock In {\em First AAAI Conference on Human Computation and
  Crowdsourcing}.
\newblock AAAI.

\bibitem[\protect\citeauthoryear{Chilton \bgroup et al\mbox.\egroup
  }{2014}]{FrenzyCHI2014}
Chilton, L.~B.; Kim, J.; Andr{\'e}, P.; Cordeiro, F.; Landay, J.~A.; Weld,
  D.~S.; Dow, S.~P.; Miller, R.~C.; and Zhang, H.
\newblock 2014.
\newblock Frenzy: Collaborative data organization for creating conference
  sessions.
\newblock In {\em Proceedings of the SIGCHI Conference on Human Factors in
  Computing Systems}, CHI '14.
\newblock New York, NY, USA: ACM.

\bibitem[\protect\citeauthoryear{Confer}{2014}]{Confer}
Confer.
\newblock 2014.
\newblock Confer.
\newblock http://confer.csail.mit.edu.

\bibitem[\protect\citeauthoryear{Confex}{2014}]{Confex}
Confex.
\newblock 2014.
\newblock Confex.
\newblock http://confex.com.

\bibitem[\protect\citeauthoryear{DiSalvo \bgroup et al\mbox.\egroup
  }{2011}]{CoDesign}
DiSalvo, C.; Lodato, T.; Fries, L.; Schechter, B.; and Barnwell, T.
\newblock 2011.
\newblock The collective articulation of issues as design practice.
\newblock {\em CoDesign} 7(3-4):185--197.

\bibitem[\protect\citeauthoryear{Ekstrand, Riedl, and
  Konstan}{2011}]{CollaborativeFiltering}
Ekstrand, M.~D.; Riedl, J.~T.; and Konstan, J.~A.
\newblock 2011.
\newblock Collaborative filtering recommender systems.
\newblock {\em Foundations and Trendstextregistered in Human-Computer
  Interaction} 4:175--243.

\bibitem[\protect\citeauthoryear{Heimerl \bgroup et al\mbox.\egroup
  }{2012}]{CommunitySourcing}
Heimerl, K.; Gawalt, B.; Chen, K.; Parikh, T.; and Hartmann, B.
\newblock 2012.
\newblock Communitysourcing: Engaging local crowds to perform expert work via
  physical kiosks.
\newblock In {\em Proceedings of the SIGCHI Conference on Human Factors in
  Computing Systems}, CHI '12,  1539--1548.
\newblock New York, NY, USA: ACM.

\bibitem[\protect\citeauthoryear{Kandasamy \bgroup et al\mbox.\egroup
  }{2012}]{CrowdDiversity}
Kandasamy, D.~M.; Curtis, K.; Fox, A.; and Patterson, D.
\newblock 2012.
\newblock Diversity within the crowd.
\newblock In {\em Proceedings of the ACM 2012 Conference on Computer Supported
  Cooperative Work Companion}, CSCW '12,  115--118.
\newblock New York, NY, USA: ACM.

\bibitem[\protect\citeauthoryear{Kim \bgroup et al\mbox.\egroup
  }{2013}]{CobiUIST2013}
Kim, J.; Zhang, H.; Andr{\'e}, P.; Chilton, L.~B.; Mackay, W.; Beaudouin-Lafon,
  M.; Miller, R.~C.; and Dow, S.~P.
\newblock 2013.
\newblock Cobi: A community-informed conference scheduling tool.
\newblock In {\em Proceedings of the 26th Annual ACM Symposium on User
  Interface Software and Technology}, UIST '13.
\newblock New York, NY, USA: ACM.

\bibitem[\protect\citeauthoryear{Kokkalis \bgroup et al\mbox.\egroup
  }{2013}]{EmailVallet}
Kokkalis, N.; K\"{o}hn, T.; Pfeiffer, C.; Chornyi, D.; Bernstein, M.~S.; and
  Klemmer, S.~R.
\newblock 2013.
\newblock Emailvalet: Managing email overload through private, accountable
  crowdsourcing.
\newblock In {\em Proceedings of the 2013 Conference on Computer Supported
  Cooperative Work}, CSCW '13,  1291--1300.
\newblock New York, NY, USA: ACM.

\bibitem[\protect\citeauthoryear{Kulkarni, Can, and
  Hartmann}{2011}]{Turkomatic}
Kulkarni, A.~P.; Can, M.; and Hartmann, B.
\newblock 2011.
\newblock Turkomatic: Automatic recursive task and workflow design for
  mechanical turk.
\newblock In {\em CHI '11 Extended Abstracts on Human Factors in Computing
  Systems}, CHI EA '11,  2053--2058.
\newblock New York, NY, USA: ACM.

\bibitem[\protect\citeauthoryear{Law and Zhang}{2011}]{CollaborativePlanning}
Law, E., and Zhang, H.
\newblock 2011.
\newblock Towards large-scale collaborative planning: Answering high-level
  search queries using human computation.
\newblock In {\em In AAAI}.

\bibitem[\protect\citeauthoryear{Macek \bgroup et al\mbox.\egroup
  }{2012}]{ConferenceAnatomy}
Macek, B.-E.; Scholz, C.; Atzmueller, M.; and Stumme, G.
\newblock 2012.
\newblock Anatomy of a conference.
\newblock In {\em Proceedings of the 23rd ACM Conference on Hypertext and
  Social Media}, HT '12,  245--254.
\newblock New York, NY, USA: ACM.

\bibitem[\protect\citeauthoryear{McNee \bgroup et al\mbox.\egroup
  }{2002}]{CitationRecommendation}
McNee, S.~M.; Albert, I.; Cosley, D.; Gopalkrishnan, P.; Lam, S.~K.; Rashid,
  A.~M.; Konstan, J.~A.; and Riedl, J.
\newblock 2002.
\newblock On the recommending of citations for research papers.
\newblock In {\em Proceedings of the 2002 ACM Conference on Computer Supported
  Cooperative Work}, CSCW '02,  116--125.
\newblock New York, NY, USA: ACM.

\bibitem[\protect\citeauthoryear{Sampson}{2004}]{AutomaticScheduling}
Sampson, S.~E.
\newblock 2004.
\newblock Practical implications of preference-based conference scheduling.
\newblock {\em Production and Operations Management} 13(3):205--215.

\bibitem[\protect\citeauthoryear{Wongchokprasitti, Brusilovsky, and
  Parra-Santander}{2010}]{ConferenceNavigator}
Wongchokprasitti, C.; Brusilovsky, P.; and Parra-Santander, D.
\newblock 2010.
\newblock Conference navigator 2.0: community-based recommendation for academic
  conferences.

\bibitem[\protect\citeauthoryear{Zhang \bgroup et al\mbox.\egroup }{2012}]{HCT}
Zhang, H.; Law, E.; Miller, R.; Gajos, K.; Parkes, D.; and Horvitz, E.
\newblock 2012.
\newblock Human computation tasks with global constraints.
\newblock In {\em Proceedings of the SIGCHI Conference on Human Factors in
  Computing Systems}, CHI '12,  217--226.
\newblock New York, NY, USA: ACM.

\end{thebibliography}
\bibliographystyle{aaai}
\end{document}